\documentstyle[amssymb,amsmath,aps,epsfig,twocolumn]{revtex}

\title{Competition of Zener and polaron phases in doped CMR manganites}
\author{ A. Wei{\ss}e~$^a$, J. Loos~$^b$, and H.~Fehske~$^a$}
\address{$^a$~Physikalisches Institut, Universit\"a{}t Bayreuth, 
  95440 Bayreuth, Germany\\
$^b$~Institute of Physics, Czech Academy of Sciences, 16200
Prague, Czech Republic\\
{\rm (\today)}\\[0.5cm]}  
\address{~\parbox{14cm}{\rm
Inspired by the strong experimental evidence for the coexistence
of localized and itinerant charge carriers close 
to the metal-insulator transition in the ferromagnetic phase 
of colossal magnetoresistive manganese perovskites, 
for a theoretical description of the  CMR transition  we propose
a two-phase scenario with percolative characteristics
between equal-density polaron and Zener band-electron states. 
We find that the subtle balance between these two states with 
distinctly different 
electronic properties can be readily influenced
by varying physical parameters, producing 
various ``colossal'' effects, such as the large   
magnetization and conductivity changes in the vicinity of 
the transition temperature.  
    }}
\begin{document}
\maketitle
\section{Introduction}
The transition from a metallic ferromagnetic low-temperature 
phase to an insulating paramagnetic high-temperature phase 
observed in some hole-doped manganese oxides (such as the perovskite family 
$\rm La_{1-x}[Sr,Ca]_xMnO_3$) is associated with an unusual 
dramatic change in their electronic and magnetic
properties. This  includes a spectacularly large negative magnetoresistive 
response to an applied magnetic field - sometimes termed 
colossal magnetoresistance (CMR) - which might have important
technological applications~\cite{JTMFRC94}. 
Starting with the  pioneer papers of Jonker and van 
Santen half a century ago~\cite{JS50},
this challenging behavior has stimulated a considerable amount of 
experimental and theoretical work~\cite{TT99,CVM99,DHM00}, 
however, even much of the basic physics of the CMR 
still remains controversial.

Early studies on lanthanum manganites concentrated on the 
link between magnetic correlations and transport, 
and attributed the low-$T$ metallic behavior to 
Zener's  {\it double-exchange mechanism}~\cite{Ze51b,AH55,Ge60}, 
which maximizes the hopping of a strongly Hund's rule coupled 
Mn $e_g$-electron in a polarized background of the 
core spins (Mn $t_{2g}$-electrons).
The quantum version of this process has been described by 
Kubo and Ohata~\cite{KO72a}.  Although there is no much controversy
about the qualitative validity of the double-exchange scenario
to stabilize a ferromagnetic state, it has been argued that physics
beyond double-exchange is important not only to explain the 
very complex phase diagram of the manganites~\cite{SRBC95,RSCCBPGBZ96} 
but also the CMR transition itself~\cite{MLS95,MYD99}. 
The difficulty is that magnetic scattering 
of itinerant charge carriers (doped holes) from enhanced 
fluctuations near $T_c$ is the exclusive mechanism to drive the 
metal-insulator transition in the double-exchange-only models. 
However, even complete spin disorder does not lead to a significant
reduction of the electronic bandwidth, and therefore 
cannot account for the observed scattering rate~\cite{Mi98,AB99,WLF01a} 
(cf. also the discussion in Refs.~\cite{FL68,ML98}). 

In view of this problem, it has been suggested that 
orbital~\cite{IYN97,KK98,HJM99,LCM00,OCP00} and, in particular, 
{\it lattice effects}~\cite{MLS95,MSM96,RZB96,ZCKM96,Baea98}  
are crucial in explaining the CMR phenomenon.   
The argument was that the electron-lattice coupling 
is known to be strong in at least some members
of the perovskite manganese family and tends to 
localize doped hole carriers as {\it small polarons} 
by changing the ratio of the energy gain through polaron formation 
to the conduction band kinetic energy~\cite{Mi98,Ki99}. 
Clearly this ratio is extremely sensitive to changes of the 
magnetic correlations by varying magnetic field, temperature 
and carrier concentration~\cite{Mi98,Ki99}.
There are two types of lattice distortions which are important 
in manganites. First the partially filled $e_g$ states of 
the $\rm Mn^{3+}$ ion are Jahn-Teller active, i.e., the
system can gain energy  from a quadrupolar symmetry elongation 
of the oxygen octahedra which lifts the $e_g$ degeneracy~\cite{St67,FS69}.
A second possible deformation is an isotropic shrinking of a  
$\rm MnO_6$ octahedron. This ``breathing''-type distortion
couples to changes in the $e_g$ charge density, i.e., 
is always associated with the presence of an $\rm Mn^{4+}$ ion.  
In the lightly doped region of the phase diagram, holes
are pinned onto $\rm Mn^{4+}$ sites by locally ``undoing'' 
the cooperative Jahn-Teller distortion, i.e., forming so-called
``anti Jahn-Teller polarons''~\cite{AP99}. 
As pointed out in a recent paper by Billinge {\it et al.}~\cite{BPPSK00}, 
this perspective is not appropriate, or at least misleading,  
in the heavily doped material, when the high-temperature 
polaronic state is approached from the ferromagnetic metallic 
(Zener) state, because there are initially almost no Jahn-Teller 
distorted octahedra. In this regime, both breathing-mode 
collapsed ($\rm Mn^{4+}$) {\it and} Jahn-Teller distorted 
($\rm Mn^{3+}$) sites are created simultaneously when the 
holes are localized in passing the metal-insulator transition.  
To avoid any confusion, in what  follows, 
we refer to  a ``polaron'' as a doped charge carrier (hole) which is  
quasi-localized with an associated lattice distortion~\cite{BPPSK00}.  
The relevance of small polaron transport above $T_c$ is obvious from
the activated behavior of the conductivity~\cite{WMG98}. 
Consequently many theoretical studies focused on polaronic
approaches~\cite{AB99,Caea97,LM97,Quea98,KJN98,Mu99,Deea99}. 
Polaronic features have been established by a variety of experiments. 
For example, high-temperature thermopower~\cite{JSRTHC96,PRCZSZ97} 
and Hall mobility measurements~\cite{JHSRDE97} 
confirmed the polaronic nature of
charge carriers in the paramagnetic phase. More directly the existence 
of polarons  has been demonstrated by atomic pair distribution~\cite{BDKNT96},
X-ray and neutron scattering 
studies~\cite{SWKT99,VROSLMSPFM99,Daea00}.
Interestingly it seems that the charge carriers partly retain their 
polaronic character well {\it below} $T_c$, as proved, e.g.,  
by neutron pair-distribution-function analysis~\cite{LEBRB97} and
very recent resistivity measurements~\cite{ZSPK00}.     
Moreover, there have been predictions, based on 
XAFS data~\cite{LSBNBRC98},  that small octahedral distortions 
persist at low temperature, forming a nonuniform metallic state. 
Particularly striking in this respect is the direct relationship 
between the structural distortions and the magnetism of the 
CMR perovskites~\cite{BBKLCN98}. That means, even the nature 
of the ferromagnetic low-temperature phase is not yet completely resolved.
Obviously manganese oxides, above and below  $T_c$, are  
in the subtle regime where many different tendencies are in 
competition and it seems that more refined ideas are needed
to explain the main properties of these materials.

Realizing that {\it intrinsic inhomogeneities} and {\it mixed-phase 
characteristics} exist and might play a key role in manganites,  
two-phase models describing the coexistence and interplay 
of localized and itinerant carriers have recently
attracted a lot of attention (for a review see Ref.~\cite{DHM00}).  
For example, {\it phase separation} scenarios involving phases with 
{\it different}  electronic densities have been adopted to describe 
the mixed-phase tendencies in manganites~\cite{MYD99,YMD98}.
These approaches are particularly meaningful    
in the limits of small ($x<0.1$) and high ($x\sim 1$) hole densities, 
where nanometer scale coexisting clusters have been reported.    
In the CMR regime ($0.15<x<0.5$), however, several experiments reported 
clusters as large as micrometers in size~\cite{FFMTAM99,UMCC99}.
Of course, such $\mu$m-sized domains, if charged, are energetically 
unstable because of the electro-neutrality condition enforced
by long-range Coulomb repulsion. Therefore an alternative concept, 
where the metal insulator transition and the associated 
magnetoresistance behaviour is viewed as a 
{\it percolation phenomenon}, was analyzed 
theoretically~\cite{GK98,GK99,MMFYD00},  
and now has a substantial experimental support by pulsed 
neutron~\cite{BE93a}, electron microscopy~\cite{UMCC99}, 
M\"ossbauer~\cite{CNIFGJG99} and scanning tunnel 
spectroscopy~\cite{FFMTAM99} studies.

Motivated by this situation, in the present work we  
propose a novel approach to the metal-insulator transition
in manganites which takes into account the percolative coexistence
of two ``intertwined''  {\it equal-density} phases:
metallic double-exchange dominated and polaronic insulating.
The transition is driven by a feedback effect which,
at $T_c$,  abruptly lowers the number of delocalized holes,
i.e., conducting sites, leading to an collapse of the bandwidth 
of the Zener state. The physical picture behind our approach
is corroborated by X-ray-absorption fine structure~\cite{BBKLCN98}
and pair distribution~\cite{BPPSK00} data, which indicate that  
charge localized and delocalized phases coexist close to the CMR
transition~\cite{BBKLCN98,BPPSK00}. Further support comes from 
zero-field muon spin relaxation and neutron spin echo 
measurements~\cite{Heea00}, reporting two time scales in the ferromagnetic 
phase of $\rm La_{1-x}Sr_xMnO_3$, where the more rapidly 
relaxing component has been  attributed to spins inside the overlapping 
metallic region and the slower Mn-ion relaxation 
rate was associated with polarons which occupy a diminishing 
volume fraction as the temperature is lowered.
\section{Two-phase model for the CMR transition}
To set up a two-component description of the metal-insulator
transition in CMR manganites, let us first characterize the 
two competing phases.
\subsection{Delocalized Zener state}
The degeneracy of $e_g$ states in (cubic) manganese oxides 
implies that the itinerant $e_g$ charge carriers carry an orbital 
degree of freedom (put in mind that the Jahn-Teller distortion lifting
the $e_g$ degeneracy vanishes in the highly-doped low-temperature
metallic phase). Consequently the $e_g$ electron transfer amplitudes 
between two Mn sites depends on the orbital orientation.
Within an $\{\theta=|3z^2-r^2\rangle,\; \epsilon=|x^2-y^2\rangle\}$ 
orbital basis, the 
anisotropic hopping matrix elements between bonds along 
the $x$, $y$, and $z$ directions are given by~\cite{SK54,KK73}   
\begin{equation}
\label{tabxy}
  t_{\alpha \beta}^{x/y} = \frac{t}{4}\left[\begin{array}{cc}
      1 & \mp\sqrt{3}\\
      \mp\sqrt{3} & 3
    \end{array}\right]\quad
  t_{\alpha \beta}^{z} = t \left[\begin{array}{cc}
      1 & 0\\
      0 & 0
    \end{array}\right]
\end{equation}
($\alpha,\beta \in \{\theta,\epsilon\}$ denote orbital 
indices). We note that the orbital pseudospin is not a conserved quantity.
The dispersion of the corresponding (noninteracting)
two-band  ($\zeta =\pm$)  tight-binding Hamiltonian is
\begin{eqnarray}
\label{ekz0}
\varepsilon_{{\bf k}\zeta}^{(0)}&=& -2 t \Big[\cos k_x +\cos k_y +\cos
k_z
\\&&\pm \Big\{\cos^2 k_x +\cos^2 k_y +\cos^2 k_z
\nonumber\\&&-\cos k_x \cos k_y
 -\cos k_y\cos k_z -\cos k_z\cos k_x\Big\}^{1/2}\Big]\nonumber\,.
\end{eqnarray}
The band structure~(\ref{ekz0}) yields the density of 
states (DOS) depicted in Fig.~\ref{fig1}
(cf. also Ref.~\cite{Ki99}). Although we employ this  DOS for all 
the calculations presented below, we would like to emphasize that  
the results will not change dramatically using an (isotropic) 
simple cubic or even constant DOS.  
\begin{figure}[!htb]
\epsfig{file= 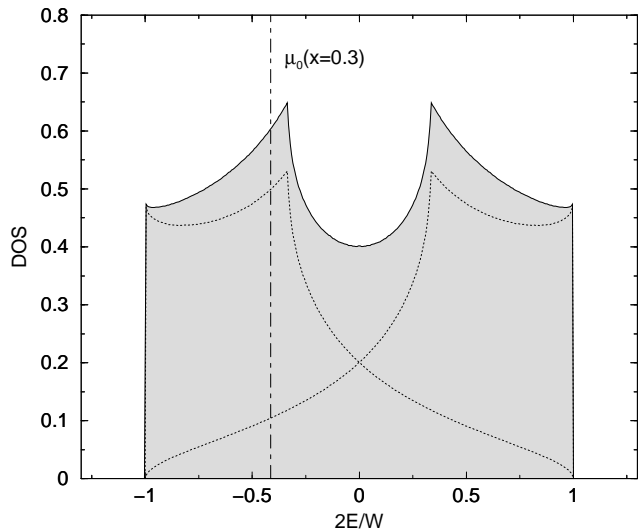, width=\linewidth}
\caption{Density of states corresponding to the band dispersion~(\ref{ekz0})
[$W=12t$ denotes the bandwidth].
Dotted lines give the DOS of the two $\zeta$-subbands. 
At optimal doping $x\simeq 0.3$, the chemical potential $\mu_0\equiv\mu(T=0)$ 
is located near the lower maximum of the DOS.}
\label{fig1}
\end{figure}

Starting from an Kondo lattice Hamiltonian, in the limits 
of strong Hund's rule coupling and large on-site Hubbard 
interaction  ($U\gg J_H\gg t_{\alpha\beta}^{x/y/z}$ which
reflects the situation in the manganites), an effective
transport Hamiltonian for (spinless) $e_g$ holes 
can be derived~\cite{KO72a,WLF01a}. 
Note that because this Hamiltonian acts in a projected 
Hilbert space without double occupied sites, in Fig.~\ref{fig1}  
the integrated DOS is normalized to one, leading
to an additional prefactor~1/2 in the $({\bf k},\zeta)$-summations. 
Then, combining the action of Zener's double-exchange  
with the percolative approach, the following effective hole 
band results
\begin{equation}
\label{ekz}
\bar{\varepsilon}_{{\bf k}\zeta}=
p^{(f)}\gamma_{\bar S}[\bar{S}\lambda]\,
\varepsilon_{{\bf k}\zeta}^{(0)}\,. 
\end{equation}
That is, in our model the renormalization of the band energy 
of the Zener state is driven by two mechanisms. 

At first, introducing an effective field 
$\lambda=\beta g\mu_BH^z_{\rm eff}$ that tends to order the 
ion spins in $z$-direction, we have the temperature- and field-dependent 
band narrowing due to the Kubo-Ohata factor~\cite{KO72a}
\begin{equation}
  \label{gsb}
  \gamma_{\bar S}[z] =  
  \mbox{\small $ \frac{1}{2}$} +\mbox{\small $ \frac{\bar S}{2\bar S + 1}$}
  \coth (\mbox{\small $\frac{2\bar S + 1}{2\bar{S}}$}z)
  \left[\coth(z) - \mbox{\small $\frac{1}{2\bar S}$} 
    \coth (\mbox{\small $\frac{z}{2\bar{S}}$})\right]\,.
\end{equation}
Here $\bar{S}=S+1/2$ refers to the total 3$d$ spin of an Mn$^{3+}$ ion
when an $e_g$ electron is present, and $S=3/2$ is the total core spin 
formed by the three $t_{2g}$ electrons. 
The factor~(\ref{gsb}), yielding an effective hole transfer amplitude
\begin{equation}
\label{td}  
\tilde{t}=\gamma_{\bar S}[\bar{S}\lambda] t\,,
\end{equation} 
was obtained averaging the transfer matrix element
for a single nearest-neighbour Mn$^{3+}$-Mn$^{4+}$ 
bond over all values and directions of the total bond spin $S_T$
(cf. Refs.~\cite{KO72a,WLF01a}). 

Next, the percolative aspects of the metal-insulator transition
discussed above imply that, at least just above $x_c$ 
(critical concentration for the occurence of the 
ferromagnetic metallic state at $T=0$) or below $T_c$, 
there exist insulating (polaronic) enclaves sparsely embedded in the 
conducting ferromagnetic (Zener) phase. Thus the ferromagnetic
phase occupies a volume smaller than the sample volume, 
i.e. $N^{(f)}<N$, where $N$ is the total number of sites
and $N^{(f)}$  denotes the number of ions in the FM phase. 
Of course the sizes and shapes of the insulating microscopic islands
are difficult to access quantitatively. 
For simplicity, within our effective two-fluid approach, we assume, 
that the hole hopping amplitude has the value $\tilde{t}$
inside the conducting region (corresponding to the effective 
transport Hamiltonian~(28) of previous work~\cite{WLF01a}) 
and zero elsewhere.
Using Fourier analysis, the homogeneous component of this
spatially dependent hopping amplitude, $\bar{t}=\tfrac{N^{(f)}}{N}\tilde{t}$,
gives the renormalization of the bandwidth, whereas the long-wavelength 
Fourier-components cause the scattering of the carriers.
Naturally the size of the ferromagnetic region, or equivalently the
``probability''  
\begin{equation}
p^{(f)}=\frac{N^{(f)}}{N}\,,
\end{equation}
has to be determined self-consistently (see Sec.~II~C). This 
introduces a {\it feedback effect}. Additional justification
for the assumed  reduction of the bandwidth proportional to
the fraction of the ferromagnetic region can be obtained 
from the numerical calculation of the density of states 
outlined in the appendix for a simple tight-binding site 
percolation model without feedback.
\subsection{Localized polaronic state}
Let us denote the number of ions and localized electron
vacancies (holes) in the polaronic phase by $N^{(p)}$ and
$N_h^{(p)}$, respectively. Of course, $N^{(p)}=N-N^{(f)}$
and $N_h^{(p)}=N_h-N_h^{(f)}$ hold, where $N_h$ is the total number 
of holes, i.e., $x=N_h/N$.  Then the energy gain due to the 
Jahn-Teller splitting on localized electron sites 
without the influence of vacancies is weakened according to    
$(N^{(p)}-N_h^{(p)})E_1=(x^{-1}-1)E_1N_h^{(p)}$, where 
$E_1$ describes an effective Jahn-Teller energy in the polaronic 
regions. At the same time, if a doped hole is localized at 
a certain site, forming a small polaron, 
a breathing distortion may occur which lowers the energy of the
unoccupied $e_g$ level by the familiar polaron
shift,  $E_p=-g^2\omega_0$ (see, e.g., Ref.~\cite{FLW00}), 
relative to its energy in an ideal structure. 
In addition, a reduction of the Jahn-Teller 
distortions in the neighborhood of the localized electron vacancy 
takes place. Both effects add up to an effective polaron binding energy 
$E_2<0 $ per hole. Neglecting the exponential small polaronic 
bandwidth, the polaronic phase, realized only in a fraction 
$p^{(p)}=N^{(p)}/N$ of the sample,   
can be represented approximately by spinless fermions (holes)
having the following site-independent energy
\begin{equation}
\label{ep}
\varepsilon_p=\left(x^{-1}-1\right) E_1 +E_2\,.
\end{equation}
As follows from a fitting of  the resistivity data to 
a small polaron model~\cite{WMG98}, 
in the region $0.2 <x<0.5$, the $x$-dependence of the 
parameter $E_2$ is rather weak, and therefore has been 
omitted in~(\ref{ep}). 
\subsection{Self-consistency equations}
The basic assumption of our mixed-phase scenario is the coexistence
of metallic and insulating clusters with equal hole density,
i.e., in accordance with recent experimental results~\cite{Heea00,BPPSK00}
(see also Ref.~\cite{MMFYD00}), we assume that there is no large-scale 
separation of $\rm Mn^{3+}$ and  $\rm Mn^{4+}$ ions in the CMR doping 
regime. This means 
\begin{equation}
\label{deftwophase}
x=\frac{N_h}{N}=\frac{N_h^{(f)}}{N^{(f)}}=\frac{N_h^{(p)}}{N^{(p)}}\,,
\end{equation}
but, of course, in general we have 
$N^{(f)}\neq N^{(p)}$ and $p^{(f)}\neq p^{(p)}$.

Denoting by $\langle{\cal H}\rangle$ the energy of the whole two-phase
system within mean-field approximation and by ${\cal S}$ the corresponding
entropy, ${\cal F}=\langle{\cal H}\rangle -T{\cal S}$ is an upper bound
for the free energy. Introducing the grand-canonical potentials
\begin{equation}
\label{omegaf}
{\mit \Omega}^{(f)} =-\frac{1}{2\beta}\sum_{{\bf k},\zeta=\pm} \ln 
\left[1+\mbox{e}^{\beta(\mu-\bar{\varepsilon}_{{\bf k}\zeta})}\right]
\end{equation}
and 
\begin{equation}
\label{omegap}
{\mit \Omega}^{(p)}=-\frac{N}{\beta}\ln 
\left[1+\mbox{e}^{\beta(\mu-\varepsilon_p)}\right]
\end{equation}
for holes in the ferromagnetic and polaronic phases, respectively, 
\begin{equation}
\label{freienerg}
{\cal F}=N_h\mu+{\mit \Omega}^{(f)} +{\mit \Omega}^{(p)} -T{\cal S}^{(s)}
\end{equation}
results, where 
\begin{eqnarray}
\label{entropie}
{\cal S}^{(s)}&=&-N\Big\{p^{(f)}\Big[(1-x)\big(
\ln \nu_{\bar S}[\bar{S}\lambda ]
- \lambda\bar{S}  B_{\bar{S}}[\bar{S}\lambda]\big)
\nonumber\\
&&\hspace*{2cm}+x \big( \ln \nu_{S}[S\lambda] -
\lambda S B_{S}[S\lambda]\big)\Big]\nonumber\\
&&\hspace*{0.7cm}+p^{(p)}\Big[
(1-x)\ln \nu_{\bar S}[0]+x
\ln \nu_{S}[0]\Big] \Big\}
\end{eqnarray}
represents the mean-field ion-spin entropy, and
\begin{eqnarray}
\label{nus}
\nu_{\bar{S}}[z]&=&\sinh (z)\coth(\tfrac{z}{2\bar{S}})+\cosh(z)\,,\\
B_{\bar{S}}[z]&=&\tfrac{2\bar{S}+1}{2\bar{S}} 
\coth (\tfrac{2\bar{S}+1}{2\bar{S}}z )-\tfrac{1}{2\bar{S}} 
\coth(\tfrac{z}{2\bar{S}})\,.  
\end{eqnarray}
If we use, instead of the model~(28), 
the spin-dependent transport Hamiltonian~(29) of our 
previous work~\cite{WLF01a}, assuming 
that the correlations of the spin background change
on a time scale large compared with the hole hopping 
frequency, i.e., we replace 
the effective hopping amplitude~(\ref{td}) by 
\begin{equation}
\label{ts}
\tilde t_{\downarrow} =  \frac{\left[\bar S (1 + B_{\bar S}
[\bar{S}\lambda])\right]^2 }{(2\bar S)(2\bar S +1)}\,t\,,
\end{equation} 
the ion-spin entropy takes the form
\begin{equation}
\label{entropie2}
{\cal S}^{(s)}\!=\!-N\Big\{p^{(f)}\big(
\ln \nu_{\bar S}[{\bar S}\lambda]- 
\lambda\bar{S}  B_{\bar{S}}[\bar{S}\lambda]\big)
+p^{(p)}\ln \nu_{\bar S}[0]\Big\}. 
\end{equation}

Most significantly, at given temperature $T$ and doping level $x$, 
{\it both} the ordering field for the ferromagnetic state ($\lambda$) and 
the size of the Zener phase ($N^{(f)}$, or alternatively the hole fraction
$N_h^{(f)}$) have to be determined in a {\it self-consistent} way, 
minimizing the free energy~(\ref{freienerg}) 
on the hyperplane $\mu(\lambda,N^{(f)})$
given by 
\begin{equation}
\label{mu}
x=\frac{1}{2N}\sum_{{\bf k}\zeta}
\frac{1}{\mbox{e}^{\beta (\bar{\varepsilon}_{{\bf k}\zeta}-\mu)}+1}
+\frac{1}{\mbox{e}^{\beta 
(\varepsilon_p-\mu)}+1}\,. 
\end{equation}
After all, the magnetization (more exact the averaged spin $z$-component) 
can be calculated from
\begin{eqnarray}
\label{M}
M&=& (1-x) \bar{S}\left[p^{(f)}B_{\bar S}[\bar{S}\lambda ]
+p^{(p)}B_{\bar S}[0]\right]
\nonumber\\
&&+ x S\left[p^{(f)}B_{S}[\bar{S}\lambda]
+p^{(p)} B_{S}[0]\right]\,. 
\end{eqnarray}
\section{Numerical results and discussion}
\begin{figure}[!htb] 
\epsfig{file=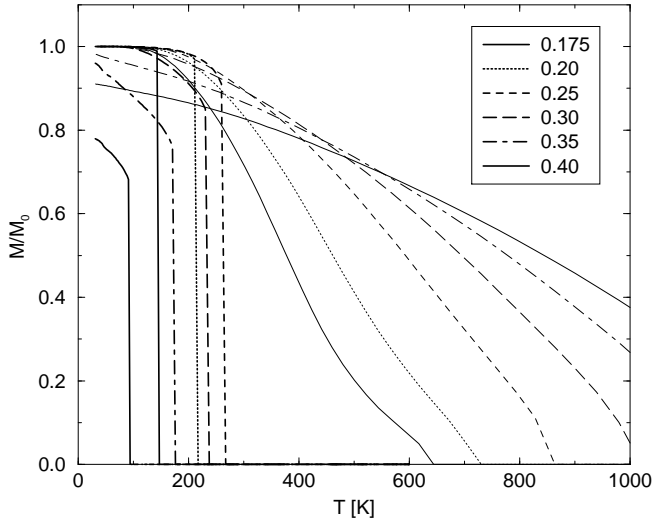, width =\linewidth}
\caption{Magnetization $M$, normalized by \mbox{$M_0\!=\!{\bar S}\!-\!x/2$}, 
as a function of temperature $T$ at various doping levels 
$x=0.175,\ldots,0.4$. 
Results are shown for the models with (bold lines) and without
(thin lines) feedback  using parameters $E_1=-0.125$~eV and 
$E_2=-0.25$~eV, and $t=0.3$~eV.}
\label{fig2}
\end{figure}

The changes in magnetization as a function of temperature 
are shown in Fig.~\ref{fig2} for different doping levels.
The efficiency of the feedback effect is quite obvious.
Omitting the $p^{(f)}$-factor in Eq.~(\ref{ekz}), a rather smooth 
variation of $M(T)$ results and the critical temperatures
for the disappearance of the Zener state, $T_c(x)$, 
are too high as compared with the experimentally observed ones.
Combining percolative and double-exchange mechanisms within our 
simple two-phase model, the magnetization exhibits a first-order 
transition, whereby $T_c$ is reduced substantially.
It can be expected that the inclusion of a small but 
finite polaron bandwidth (instead of the localized level~(\ref{ep})),   
softens the abrupt transition to some extent  but in any case we
will find a rather sharp drop of  $M(T)$ near $T_c$.

\begin{figure}[!htb] 
\epsfig{file= 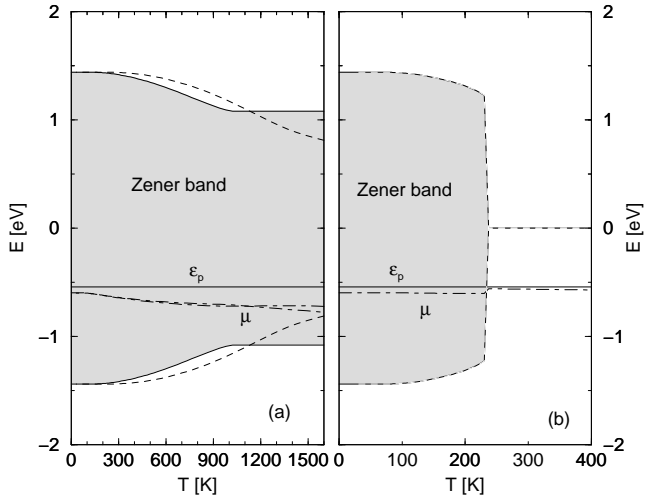, width =\linewidth}
\caption{Temperature-dependences of the Zener band (shaded region)
and of the  positions of the polaronic level ($\varepsilon_p$) and 
chemical potential ($\mu$) without (a) and with (b) feedback.
Results are presented at $x=0.3$ ($E_1=-0.125$~eV, $E_2=-0.25$~eV; 
$t=0.3$~eV).
Dashed lines display the band edges resulting from~(\ref{ts})
instead of~(\ref{td}).}
\label{fig3}
\end{figure}

Naturally the bandwidth of the Zener state, displayed
in Fig.~\ref{fig3} for a doping level $x=0.3$, reflects 
the behavior of the magnetization. Without feedback we found
a moderate band narrowing, whereas a radical shrinking occurs 
in the feedback model, which can be traced back to the collapse 
of the ferromagnetic metallic region at $T_c$.
The use of the spin-dependent hopping amplitude~(\ref{ts})
leads to a modification of the Zener bandwidth (and $\mu$)  
only for the scenario without feedback. Here, 
temperature-independent band edges correspond to 
complete spin disorder. Including the $p^{(f)}$-factor, 
both models~(\ref{ts}) and~(\ref{td}) yield 
nearly identical results, which again indicates the 
dominance of the percolative feedback mechanism.
\begin{figure}[!htb] 
\epsfig{file= 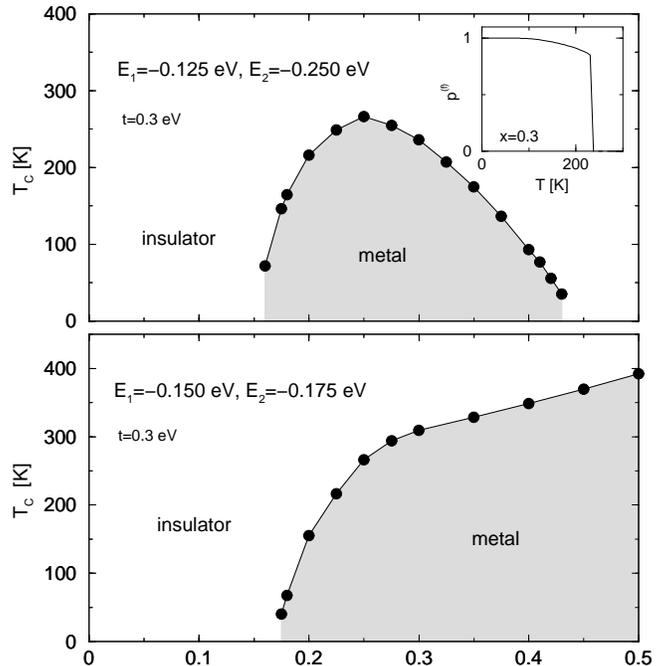, width = \linewidth}
\caption{Phase diagram of the mixed-phase Zener-polaron model with feedback.
The inset shows the fraction of the Zener phase as a function of temperature.}
\label{fig4}
\end{figure}
Figure~4 presents the central result of our work, 
the $x$-$T$ phase diagram of the percolative two-phase model.
First of all we would like to state that it is clearly beyond
the scope of the present approach to account for the complete 
phase diagram of real manganites, which contains a great 
variety of charge, spin and orbital ordered states.     
The focus is on the doping region $0.15<x<0.5$, 
where at low temperatures a band-description seems to be 
appropriate.  In this regime, the phase diagrams of Fig.~4,
determined for two characteristic  parameter sets,  
($E_1=-0.125$~eV, $E_2=-0.25$~eV, $t=0.3$~eV) and  
($E_1=-0.15$~eV, $E_2=-0.175$~eV, $t=0.3$~eV),
describe the major features of the metal-insulator transition
lines in $\rm La_{1-x}Ca_xMnO_3$ and $\rm La_{1-x}Sr_xMnO_3$~\cite{IFT98}.     
Furthermore, even the absolute values of the critical concentration 
$x_c(T=0)\simeq 0.17$ and of the transition temperatures $T_c$
agree surprisingly well with the experimental data.
 
As can be seen from the inset, the low-temperature phase 
contains besides the ferromagnetic metallic region finite
domains of the localized polaronic phase ($p^{(f)}<1$,
and accordingly $p^{(p)}>0$), the fraction of which increases 
with increasing temperature. Finally, at $T_c$, there is an 
abrupt spill-over of holes between the delocalized and localized phases,
which drives the metal insulator transition. Since the conductivity 
is proportional to the number of delocalized holes $N_h^{(f)}$,
a dramatic change at $T_c$ results. In the insulating 
high-temperature phase, the band description
breaks down, and the transport properties are dominated by 
incoherent small polaron hopping processes. For doping levels
$x<x_c$, the system behaves as a short-range correlated Jahn-Teller
insulator. 
\section{Summary}
Although the theoretical understanding of the 
CMR phenomenon is still incomplete, double-exchange, electron-phonon 
and orbital effects are commonly accepted as the main ingredients.
Based on an in-depth analysis of important new experimental
information about the inhomogeneous microscopic structure 
of the ferromagnetic metallic state, we come to the conclusion 
that in addition mixed-phase tendencies and percolative behavior play an
important role in doped manganites, in particular in the vicinity 
of the metal-insulator transition. To simulate these effects,
we have studied a simple semi-phenomenological model,
describing coexisting polaronic insulating and double-exchange dominated 
metallic phases with equal hole density. The fraction of localized 
and delocalized states was determined self-consistently.
Below the transition temperatue $T_c$, we found polaronic
inclusions embedded in a dominant macroscopic metallic phase. 
Our approach, which is distinct from that of the electronic 
phase separation concept involving states with different carrier
density, provides an alternative explanation 
of the ferromagnetic metal to polaronic insulator transition.
The abrupt change, revealed in various electrical and magnetic
properties at $T_c$, was attributed to a collapse of the Zener state 
mainly caused by a percolative feedback mechanism.  
At $T=0$ the transition is driven by doping and occurs
at $x_c\simeq 0.15-0.18$. At finite temperatures, 
disorder due to intrinsic inhomogeneities and magnetic scattering 
act in  combination to reduce the mobility of the charge carriers. 
The calculated values of $T_c$ agree fairly well with the experimental ones.
In conclusion, we believe that a further elaboration of the proposed 
percolative mixed-phase scenario, e.g., towards a direct calculation of the 
transport properties, may help to clarify many puzzling 
aspects of the CMR compounds. 
\section*{Acknowledgements}
The authors are greatly indebted to L. F. Feiner and R. Kilian
for stimulating discussions.
This work was supported by the Deutsche Forschungsgemeinschaft
and the Czech Academy of Sciences under Grant No. 436 TSE 113/33.
\section*{Appendix: Percolative picture}
To support the assumption that the bandwidth of the 
Zener state depends approximately linear on the fraction of 
the ferromagnetic region (see Sec.~II~B, Eq.~(\ref{ekz})), 
let us consider a site percolation model 
on a finite hypercubic lattice with 64$^3$ sites 
and periodic boundary conditions. The lattice points 
are occupied with probability $p$. Adjacent
occupied sites will be connected by a hopping matrix
element. The density of states 
of the resulting random tight-binding model,
\begin{equation}
{\cal H}_{p}=\sum_{\langle i j\rangle}t_{p} 
(c_i^{\dagger}c_j^{\dagger}+c_j^{\dagger}c_i^{\dagger})
\end{equation} 
$t_p\in \{0,1\}$, 
is determined numerically in two different ways,
using kernel polynomial and maximum entropy methods~\cite{SRVK96,BWF98}
On the one hand, we start from a single realization ${\cal H}_p$ 
and determine 400 Chebyshev moments of ${\cal H}_p$ using 100 random 
start vectors. On the other hand, we use 10 realizations of ${\cal H}_p$
and compute for each realization 200 moments from 50 random 
start vectors. The results basically agree. 
\begin{figure}[!htb] 
\epsfig{file= 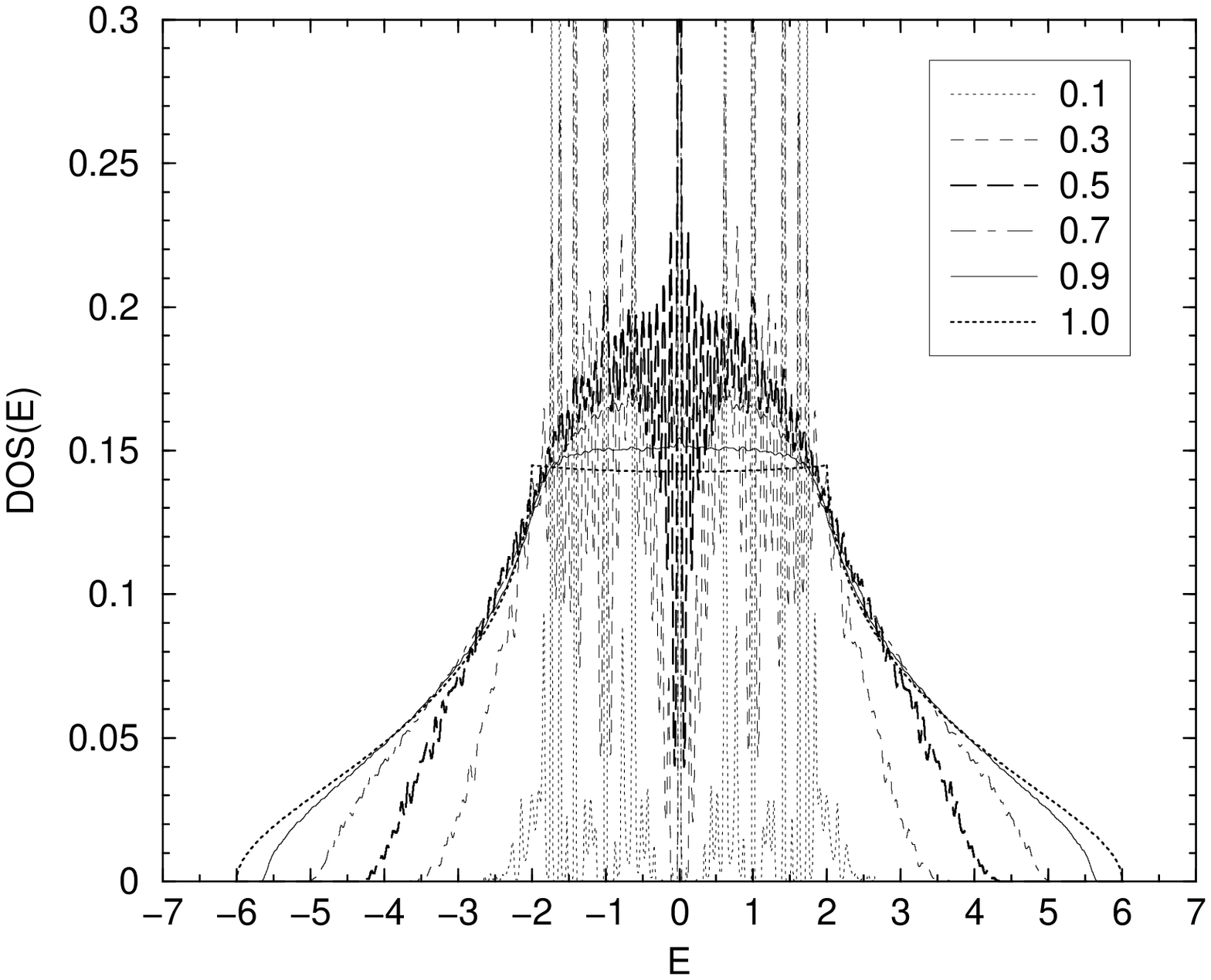, width =0.98\linewidth}
\hspace*{0.3cm}\epsfig{file= 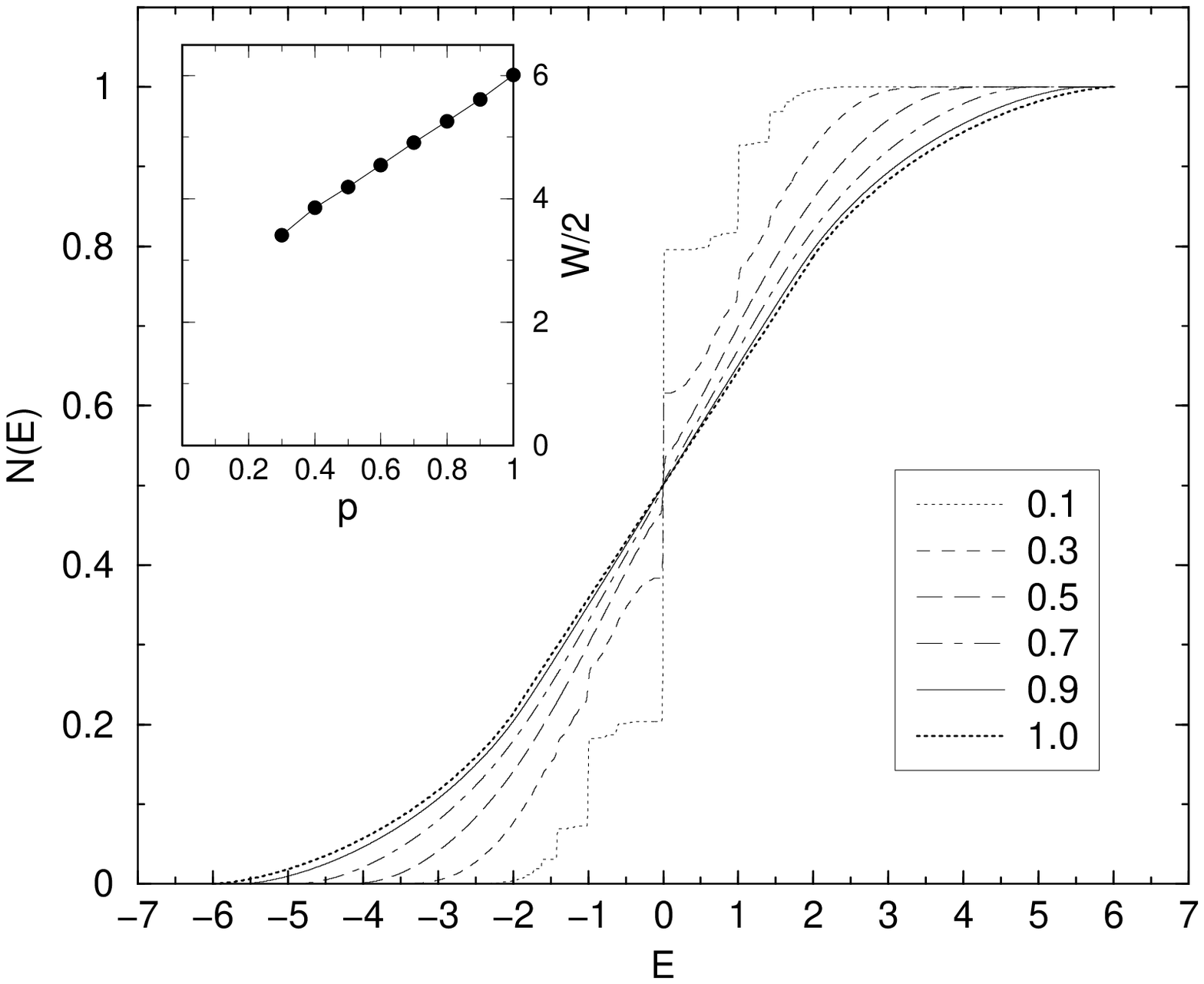, width =0.98\linewidth}
\caption{Density of states (DOS) (upper panel) and integrated DOS ($N(E)$)
for the tight-binding site percolation model with different probabilities 
$p=0.1, 0.3,\ldots ,1$. The inset gives the bandwidth as a function of $p$.}
\label{fig6}
\end{figure}

Figure~\ref{fig6} displays the (averaged) density of states 
for various values of $p$. Self-evidently, $p$ now is simply  
a model parameter, i.e., there is no feedback effect as considered 
in Sec.~II~B. A significant change of the spectrum around the 
site percolation threshold, $p_c\simeq 0.31$, is observed.
Below  $p_c$, the DOS is mainly confined to the interval [-2,2],
like for a one-dimensional model, and mostly consists of isolated
peaks with different spectral weight. This can be seen more clearly
from the steplike behaviour of the integrated density of states, 
$N(E)=\int_{-W/2}^E {\rm DOS}(E') dE'$, shown in the lower panel.
The steps can be attributed to metallic ``islands'' (finite clusters) 
containing delocalized electrons. They form a cluster comparable to
the system size first at $p=p_c$. Above $p_c$, $N(E)$ becomes a
continuous function and the DOS is evocative of that of 
a simple cubic tight-binding model but with a reduced bandwidth.
That is, although the percolative cluster occupies most of the 
bulk, making the whole system metallic, smaller conducting and insulating 
regions are embedded into it. These enclaves act as small scattering centers,  
causing the band narrowing. The inset demonstrates the almost perfect 
linear dependence of $W$ on $p$, in particular at large $p$, which 
in reality is the region of interest (cf. the inset of Fig.~\ref{fig4}).

\end{document}